\documentclass[12pt,a4paper]{article}

\usepackage{amssymb,amsmath,bbm}
\usepackage{multirow}
\usepackage{amsthm}
\usepackage{upgreek}
\usepackage{graphicx}
\usepackage{float}
\usepackage{authblk}
\usepackage{color}
\usepackage{soul}
\usepackage{subcaption}

\newtheorem{theorem}{Theorem}
\newtheorem{definition}{Definition}
\newtheorem{assumption}{Assumption}

\begin{document}

	\title{An Efficient Unified Approach for Spread Option Pricing in a Copula Market Model}
	\author[1]{Edoardo Berton}
	\author[2]{Lorenzo Mercuri}
	\affil[1,2]{Department of Economics Management and Quantitative Methods,
		University of Milan.}
	\affil[2]{CREST Japan Science and Technology Agency.}
	\maketitle
	
	\begin{abstract}
		In this study, we propose a new formula for spread option pricing with the dependence of two assets described by a copula function. The advantage of the proposed method is that it requires only the numerical evaluation of a one-dimensional integral.
		Any univariate stock price process, admitting an affine characteristic function, can be used in our formula to get an efficient numerical procedure for computing spread option prices.\newline In the numerical analysis we present a comparison with the Monte Carlo simulation method to assess the performance of our approach, assuming that the  univariate stock price follows three widely applied models: Variance Gamma, Heston's Stochastic Volatility and Affine Heston Nandi GARCH(1,1) models. 		
		\bigskip\newline
		\textbf{keywords}: Copula Function, Affine process, Spread Options
	\end{abstract}
	
	\section{Introduction}
	\label{section:INTRO}
	
	A spread option is a contract written on the price difference (spread) between two assets. For example, the European call spread option gives the holder the right to receive this difference by paying a strike price at maturity.
	
	The most common modelling approach in the literature, as well as the prevailing in practice, is the multivariate geometric Brownian motion (henceforth MGBM) approach. MGBM model is often used alongside Monte Carlo simulation and provides the framework for further developments. The approach relies on the linear correlation coefficient to collect full information about the processes’ dependence structure. Many authors have provided their contribution to this approach by deriving a closed form solution to the pricing problem. Notably, \cite{margrabe} derives an option pricing formula when the exercise price is zero. \cite{kirk} provides a further approximation that is currently the market standard for the contract. \cite{caramona} compute an analytical approximation by reducing the price range to a bounded interval that is particularly tight for certain parameters values. Finally, \cite{BSSpread} introduce a lower bound for the price that has been proven to be more accurate than the approximation in \cite{kirk}.
	
	The vivacity of the market for spread options has recently drawn much attention. Several works have provided new theoretical insights on the topic of spread options. \cite{caldana_fusai} and \cite{vanduffel} present new methodologies for the computation of the price of spread options, while \cite{malliavin} provides an assessment of spread option price sensitivities through Malliavin calculus. Further enrichment to the spread option framework are introduced by \cite{benth} and \cite{schneider}. The former proposes a joint model for day-ahead electricity prices in interconnected markets and observes that the model yields analytical prices. The latter develops a multi-factor model and shows how to compute the joint characteristic function of a pair of futures prices.
	
	An alternative method to price spread options by means of copula functions is provided in \cite{chiou} and \cite{HHK}, with copula-based dependence structure imposed on the daily GARCH innovations. These works propose a copula-based dependence structure and different processes for underlying assets. Previous literature exhibits two major limitations: (i) past pricing methods rely heavily on Gaussian framework, even though different processes may better reflect the underlying's behaviour, and (ii) most of the literature (although not all) imposes linear correlation as a measure of dependence. 
	
	In this study, we develop a framework for spread option pricing that exploits the flexibility of the copula function in capturing the dependence between the marginals without loss of fit on other crucial features of financial data (\textit{e.g.} heavier tails, non linear dependence and asymmetrical dependence). Considering an appropriate copula family, we derive a formula for a spread option that involves the first partial derivatives of the copula function. 
	
	The proposed method requires only the cumulative distribution function of marginals and the corresponding quantile functions.These quantities are easy to compute for most continuous and discrete time market models. 
	The most popular continuous time models can be classified into two groups: L\'evy processes and stochastic volatility models. The main difference between two groups is the characterization of the process used for the increments. In particular, L\'evy processes have stationary and independent increments with a constant volatility while, for the latter group, the volatility is driven by a specific stochastic differential equation with a second source of randomness. 
	
	The distribution at time one of a L\'evy process is infinitely divisible; its characteristic function is obtained using the L\'evy–Khintchine formula and, for all L\'evy processes, the cumulative distribution function can be obtained numerically by the Gil-Pelaez inversion formula in \cite{gilpelaez}.  L\'evy processes provide good fitting to asymmetry and fat-tails actually observed in financial time series. We mention a special subgroup of L\'evy processes named Time-Changed Brownian motion, that arise by considering a Brownian motion evaluated at a random time represented by a subordinator L\'evy process.  Time-Changed Brownian motion has become popular in modelling financial prices mainly due to the fact that the subordinator is able to reproduce the market time activity \cite{clark1973subordinated}. The distribution at time one is a Normal Variance Mean Mixture \cite{barndorff1982normal} and it has as special cases many models considered in financial literature such as the Variance Gamma \cite{madan1990variance}, the Normal Inverse Gaussian \cite{barndorff1982normal} and the Generalized Hyperbolic \cite{eberlein2002generalized} distributions. The cumulative distribution function of these distributions can be efficiently approximated with a finite mixture of normals as an alternative to the inversion formula in \cite{gilpelaez}, see \cite{loregian2012approximation,mercuri2021finite} for the Variance Gamma and a more generic case, respectively. Other possible L\'evy processes that can be used in our framework are the Tempered Stable \cite{carr2003stochastic}, the  Mixed Tempered Stable \cite{rroji2015mixed,mercuri2018option} and the $\alpha$-Stable with maximum negative skewness \cite{carr2003finite}. 
	
	Among the continuous-time stochastic volatility models that can  be applied in our context for the determination of the marginal cumulative distribution function, the most relevant are those that imply an affine structure for the log-price conditional characteristic function. Indeed, the characteristic function has a log-linear affine structure with respect the current level of variance with time-dependent coefficients that are solutions of a system of ordinary differential equations. Starting from the seminal work \cite{heston}, we have observed an increasing interest on the latter models. In particular, a unified framework for affine asset pricing models in continuous-time  has been proposed in \cite{duffie2000transform}, while \cite{barndorff2001non} considers a stochastic volatility model where the variance process is an Ornstein-Ulhenbeck driven by a L\'evy subordinator. See \cite{muhle2012option} for its affine structure. Recently, affine stochastic volatility models  with an additional source of risk have been introduced in \cite{bernis2021gamma, brignone2020asian}. This new component is a compound Hawkes process that produces a self-exciting effect of the large jump in the log-price dynamics. 
	
	In discrete-time, among the most relevant processes for financial log-returns are the GARCH models introduced in \cite{Bollerslev} as a generalization of the ARCH model presented in \cite{engel}. GARCH models assume that the volatility process at time $t$ is influenced not only by previous error terms but also by its lagged values, favouring volatility clustering dynamics. Another advantage of GARCH modelling is that it does not require returns to be independent. Indeed, despite being considered Gaussian distributed conditional on past values, returns' unconditional distribution exhibits leptokurtosis. 
	
	
	From the seminal work in \cite{duan1995}, several GARCH models for option pricing have been proposed in the financial literature. A major breakthrough occurred with the GARCH model with Gaussian innovations developed in \cite{HNGARCH} where the specification of the variance process yields a recursive procedure for the computation of the characteristic function of the log price. To distinguish this model from to the original GARCH, the model in \cite{HNGARCH} is known in the literature as affine GARCH model with Gaussian innovation.  Affine GARCH model with non-Gaussian innovations are the model proposed in \cite{ig_garch} with Inverse Gaussian innovations and the model in \cite{MERCURI2008172} with Tempered Stable innovations. 
	
	For completeness, it is worth to notice that our procedure can be combined with non-parametric methods for estimating the marginal risk neutral density from option prices as proposed in \cite{breeden1978prices}.
	
	Although all the aforementioned models can be included in our framework for spread option pricing, we present a numerical comparison by considering only one model in each group discussed above. Indeed, we employ the Variance Gamma \cite{madan1990variance}, the Heston continuous-time stochastic volatility model \cite{heston} and the affine GARCH model with normal innovations \cite{HNGARCH}. To conduct a fair comparison among these models the cumulative distribution function has been obtained by means of the inversion formula \cite{gilpelaez}. Numerical results of our formula are benchmarked against an adequate Monte Carlo simulation to assess the accuracy and the performance of the introduced method.
	 
	This paper is organized as follows. Section \ref{section:COP} reviews the main properties of copula functions. We propose a numerical method for pricing spread options based on the computation of a single integral in Section \ref{SOPC} while, in Section \ref{section:AGM}, we review the main features of the three univariate option pricing models that we use to test the performance of our procedure. Indeed, we analyze numerically the precision of the formula in Section \ref{Num}. In Section \ref{Sect:CopGARCH}, we remark that, contrary to the procedure developed in \cite{chiou}, the dependence structure of log-returns is described by a copula function on the entire time to maturity window and, therefore, we discuss a procedure for converting the parameters of our pricing formula to those of a bivariate GARCH process where the innovations are correlated according to a copula function. Section \ref{conclude} concludes the paper.
	
\section{Copula Function and Dependence Structure}\label{section:COP}
In this section we briefly report the theoretical concept of bivariate copula, outlining few principal results required in the following of the paper. We describe shortly the Archimedean copulas and summarize the copula functions that are used at a later step, specifically Gumbel, Clayton, Frank and Plackett copula. These copulas have been widely applied in finance (\cite{bellini2020dependence,chiou,HHK,rosenberg2003non}).




First, let us recall the basic definition of a bivariate copula function. A bivariate
copula $C: [0,1]^2 \rightarrow [0,1]$ is the distribution function of a two-dimensional random vector with uniform marginals. According to Sklar’s theorem, given a random vector $(X_1,X_2)$ with joint distribution function $F(x_1, x_2) = P(X_1 \leq x_1, X_2 \leq x_2)$ and with continuous marginals’ distributions $F_1, F_2:\mathbb{R} \rightarrow [0,1]$, there exists a unique copula $C$ such that
\begin{equation}\label{sklar theorem}
	C(F_1(x_1),F_2(x_2)) = F(x_2, x_2)\;\;\;\;\;\;\;\; x_1,x_2\in \mathbb{R}
\end{equation}
An interesting implication of Sklar's theorem is that, conditional on the marginal distributions $F_1$ and $F_2$ being absolutely continuous, the corresponding unique copula distribution can be rewritten as
\begin{equation}
	C(u, v) = F(F_1^{-1}(u), F_2^{-1}(v))
\end{equation}
The latter illustrates how copulas can be extracted from multivariate joint cumulative distribution functions with continuous margins. Noting that $C(u,v)$ corresponds to the probability that $X_1$ and $X_2$ lie below the $u$-th and $v$-th quantile respectively, it appears clear how copulas express dependence on a quantile scale.

It seems appropriate to define how the copula function is distributed conditional on the marginals.
\begin{definition}
	The conditional distribution function of the copula $C(U,V)$ given $V = v$, which we denote $c_v(u)$, is given by:
	\begin{equation}\label{eq:cond_distribution}
		c_v(u) = \lim_{\Delta v \rightarrow 0}\frac{C(u, v + \Delta v) - C(u,v)}{\Delta v} = \frac{\partial C(u,v)}{\partial v}
	\end{equation}
\end{definition}
The conditional distribution $c_v(u)$ is a distribution in $\left[0,1\right]$ and is uniform if and only if the copula corresponds to the independence copula $\Pi$. The conditional distribution function is particularly useful when simulating a random sample that is jointly distributed according to a copula $C$.

The second order mixed partial derivative of a copula function provides the joint density function, denoted $c(u,v)$. Formally, in the bivariate case
\begin{equation}\label{eq:copula_density}
	c(u,v) = \frac{\partial^2 C(u,v)}{\partial u \partial v}
\end{equation} 

Not all copulas possess, however, a closed form density function. For instance, the countermonotonicity copula and comonotonicity copula are not absolutely continuous and do not admit the computation of the density function. Yet, most parametric specifications provide analytical densities to ease further computations.

Copula functions can be explicitly defined or extracted from an existing joint distribution by virtue of Sklar's theorem results. However, it is often the case that the latter cannot be extracted in a simple closed form.

In contrast with implicit copulas, explicitly defined copulas arise from the need of a simple closed-form algebraic expression to compute joint probability. Among these explicit copulas, Archimedean copulas enjoy considerable popularity in a number of practical applications in several disciplines. This collection of copulas has in fact a convenient algebraic expression and is capable of capturing asymmetry.
\begin{definition}
	A copula $C_{\varphi}$ is an Archimedean copula if it possesses the functional form
	\begin{equation}\label{archimedean copula}
		C_{\varphi}(u, v) = \varphi(\varphi^{-1}(u),\varphi^{-1}(v))
	\end{equation}
	For a suitable non-increasing function $\varphi:[0,\infty) \rightarrow [0,1]$ with $\varphi(0)=1$ and $\lim_{x \rightarrow \infty} \varphi(x) = 0$, called the ``Archimedean generator''.
\end{definition}
\cite{Malov} postulates that $C_\varphi$ is a valid distribution if the Archimedean generator $\varphi$ is $d$-monotone. The generator $\varphi$ is said to be \textit{d-monotone} if it has derivatives of all orders in $(a,b)$ and if $(-1)^k f^{(k)}(x) \geq 0$ for any $x \in(a,b)$ and for $k=0,1,...,d$.
Each completely monotone generator function $\varphi$ can be applied to define a copula of arbitrary dimension $d\geq2$. \newline
Although Archimedean copulas are typically parametrized in terms of a function $\varphi$, in practice one chooses a parametric family of Laplace transforms governed by some real parameter $\theta$, so that $\varphi = \varphi_\theta$. 
This allows the copula to be completely parametrized by the single parameter $\theta$, so that  $C_\theta$ is equivalent to $C_{\varphi_\theta}$. Hence, it is the parameter $\theta$ that provides the range of different dependence structures. \newline	
Table \ref{tab:archimedeancop} shows three Archimedean copulas and the respective generator function. We exploit Clayton, Gumbel and Frank copulas in the empirical exercise in order to benchmark the accuracy of the novel semi-analytical spread option pricing technique.

\begin{table}[]
	\centering
	\begin{tabular}{ c c c c}
		\hline \\[-10pt]
		\small name & \small $\varphi_\theta (x)$ & $C(u,v)$
		& \small $\theta \in$ \\
		\hline \\ [-8pt]
		\small Clayton copula & \small $(1+x)^{-1/\theta}$ & \small$(u^{-\theta} + v^{-\theta} -1)^{-1/\theta}$ &\small $(0,\infty)$ \\[9pt] 
		\small Gumbel copula & \small $\exp\{-x^{1/\theta}\}$ & \small$\exp(-((-\log u)^\theta + (-\log v)^\theta)^{1/\theta})$ &\small $[1,\infty)$ \\[9pt] 
		\small Frank copula  & \small $-\log\left(\frac{e^{-\theta x} -1}{e^{-\theta} -1}\right)$  &\small$-\frac{1}{\theta}\log\left(1+ \frac{(e^{-\theta u} -1)(e^{-\theta v} -1)}{e^{-\theta} -1}\right)$ & \small$(-\infty, \infty)\backslash\{0\}$\\[4pt]
		\hline
	\end{tabular}
	\caption{\label{tab:archimedeancop}Archimedean copulas with the respective generator function $\varphi$ and $\theta$ parameter range.}
\end{table}

In addition to the three Archimedean copulas we mentioned above, the empirical analysis rests heavily on the use of Plackett copula. It seems then appropriate to briefly report some results about this distribution.

This bivariate distribution, introduced in \cite{plackett}, provides an extension of the concept of contingency table dependence, allowing for continuous margins. The copula has been proved to be ``comprehensive'' in that it allows modelling of independence as well as positive and negative dependence structures (\textit{i.e.} it ranges continuously from the lower to the upper Fréchet–Hoeffding bound).

The analytical expression of the Plackett copula reads 
\begin{subequations}
	\begin{equation}\label{plackett copula1}
		\hfilneg 
		C(u,v) = \frac{[1+(\theta-1)(u+v)] - \sqrt{[1+(\theta-1)(u+v)]^2 - 4uv\theta(\theta-1)}}{2(\theta-1)}
		\hspace{10000pt minus 1fil}
	\end{equation}
	for $\theta > 0$ and $\neq 1$.
	\begin{equation}\label{plackett copula2}
		C(u,v) = uv
	\end{equation}
	for $\theta=1$.
\end{subequations}

By taking the mixed second order partial derivative of $C(u,v)$ we obtain the following closed-form analytical probability density function 
\begin{equation}\label{plackett density}
	c(u,v) = \frac{\theta\left[1 + (u-2uv+v)(\theta-1) \right]}{\{ \left[1+(\theta-1)(u+v) \right]^2 - 4uv\theta(\theta-1) \}^{3/2}}
\end{equation} 

The Plackett copula lends itself particularly well to modelling dependence structure of financial phenomena, and it is well spread in practice. It possesses analytical first and second order partial derivatives and, as only one parameter $\theta$ captures the entire dependence dynamic, it is notably easy to estimate via Maximum Likelihood from past data. One could also estimate $\theta$ by computing the observed cross-product ratio in the dataset using some arbitrary quadrant separation. 

\begin{figure}[h]
	\centerline{\includegraphics[width=1.3\textwidth]{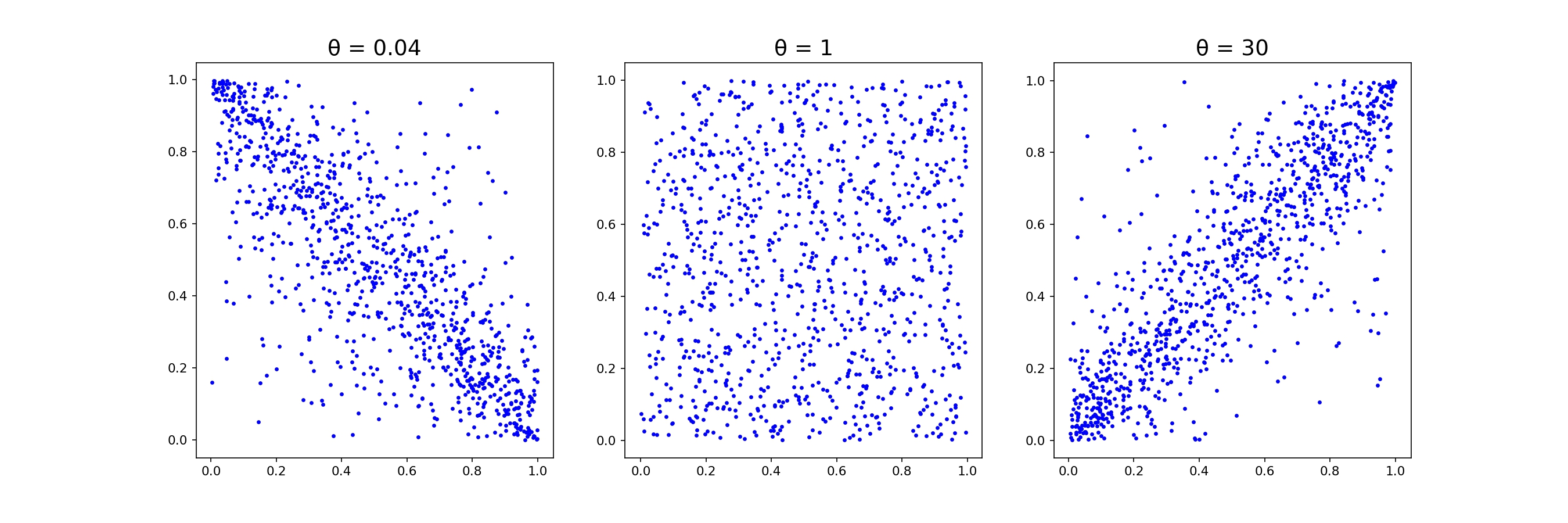}}
	\caption{Scatter plot of Plackett copula in the following cases: negative dependence ($\theta <1$), independence ($\theta=1$) and positive dependence ($\theta >1$).}
	\label{fig:plackett_scatter}
\end{figure}

The scatter plots in Figure \ref{fig:plackett_scatter} show the pattern in the joint distribution, when $\theta$ ranges from $\theta=0.04$ to $\theta=30$. The two lateral plots exhibit clear negative and positive dependence respectively, whilst the middle one represents independence.

\section{Spread Option Pricing with a Copula Function}
\label{SOPC}

In this section we propose a method for pricing a spread option based on a numerical integration of one-dimensional integral. We assume that the joint distribution of log-returns is modeled through  a copula function. We discuss only the case of a European spread call option while the case of a European spread put option is obtained by using the standard put call/parity. The formula proposed in this section is easy to apply to any model where the univariate cumulative distribution function has an analytical (or semi-analytical) formula. Under a martingale measure $\mathbb{Q}$, the price of a spread call option with strike price $K$ can be obtained as:
\begin{equation}
	P_{t_0}=e^{-r\left(T-t_0\right)}\mathsf{E}^{\mathbb{Q}}_{t_0}\left[\max\left\{S_{1,T}-S_{2,T}-K;0\right\}\right].
	\label{eq:ref1}
\end{equation}

Denote  by $\left(\Omega,\mathcal{F},\left(\mathcal{F}_t\right)_{t\geq0},\mathbb{P}\right)$ the underlying complete filtered space. 
Defining with $X_{i,t_0,T}:=\log\left(S_{i,t}\right)-\log\left(S_{i,t_0}\right)$ with $i=1,2$ the log-returns over the time to maturity interval $\left[t_0,T\right]$, we report here the notations and conventions used in this section. 

\begin{itemize}
	\item For the conditional distribution of $X_{i,t_0,T}$, we write $\forall T>t_0$ 
	\begin{equation*}
		F_{i,t_0,T}\left(x\right):=\text{Prob}\left[\left\{X_{i,t_0,T} \leq x\right\} \left|\mathcal{F}_{t_0} \right.\right].
	\end{equation*}
	\item The function $f_{i,t_0,T}\left(x\right)$ stands for the conditional density of $X_{i,t_0,T}$ (i.e. $f_{i,t_0,T}\left(x\right):= \frac{\partial F_{i,t_0,T}\left(x\right)}{\partial x}$).
	\item For the conditional quantile function of $X_{i,t_0,T}$, we write $q_{i,t_0,T}\left(u\right)$ that is $q_{i,t_0,T}\left(F_{i,t_0,T}\left(x\right)\right)=x$. We define  with $\tilde{q}_{i,t_0,T}\left(u\right):=\exp\left[q_{i,t_0,T}\left(u\right)\right]$. Due to the fact that conditional distribution $F_{i,t_0,T}\left(x\right)$ is a strictly increasing function, $q_{i,t_0,T}\left(u\right)$ and $\tilde{q}_{i,t_0,T}\left(u\right)$ are both strictly increasing functions.
	\item For the joint conditional cumulative distribution function of the couple $\left(X_{1,t_0,T},X_{2,t_0, T}\right)$ we write:
	\begin{equation*}
		F_{t_0,T}\left(x_{1},x_{2}\right):=\text{Prob}\left[\left\{X_{1,t_0,T} \leq x_{1}\right\} \bigcap \left\{X_{2,t_0,T} \leq x_{2}\right\} \left|\mathcal{F}_{t_0} \right.\right]
	\end{equation*}
	and the associated joint conditional density is denoted by $f_{t_0,T}\left(x_{1},x_{2}\right)$, \textit{i.e.}
	$f_{t_0,T}\left(x_{1},x_{2}\right)=\frac{\partial^2 F_{t_0,T}\left(x_{1},x_{2}\right)}{\partial x_{1},\partial x_{2}}, \ \forall T>t_0$.
	\item For  the first order partial derivatives of the copula function $C\left(u,v\right)$, we write $c_{v}\left(u\right):=\frac{\partial C\left(u,v\right)}{\partial v}$ and $c_{u}\left(v\right):=\frac{\partial C\left(u,v\right)}{\partial u}$ while the mixed second order partial derivative is denoted by $c\left(u,v\right)= \frac{\partial^2 C\left(u,v\right)}{\partial u \partial v}$. 
\end{itemize}

Using the copula function $C\left(u,v\right)$, the joint density $f_{t_0,T}\left(x_{1},x_{2}\right)$ of log-returns  can be written with our notation in the following way: 
\begin{equation*}
	f_{t_0,T}\left(x_{1},x_{2}\right)=c\left(F_{1,t_0,T}\left(x_{1}\right),F_{2,t_0,T}\left(x_{2}\right)\right)f_{1,t_0,T}\left(x_{1}\right)f_{2,t_0,T}\left(x_{2}\right)
\end{equation*}
For any twice differentiable copula function, the formula in \eqref{eq:ref1} can be obtained using the following result:
\begin{theorem} \label{MainTheorem}
	
	Under the equivalent martingale measure $\mathbb{Q}$, we consider a market composed of two risky assets such that, for any $T>t_0$, both prices satisfy:
	\[
	S_{1,T}=S_{1,t_0}e^{X_{1,t_0,T}}
	\]
	and
	\[
	S_{2,T}=S_{2,t_0}e^{X_{2,t_0,T}}
	\]
	where $X_{i,t_0,T}$ with $i=1,2$ are the log-returns over the interval $\left[t_0,T\right]$ and we assume $r$ to be the continuously compounded risk-free interest rate.\newline
	If the joint distribution of $(X_{1, t_0, T}, X_{2, t_0, T})$ is represented by a twice differentiable copula (i.e. $C\left(u,v\right) \in \mathcal{C}^{2,2}$), the no-arbitrage price for the spread call option is:
	\begin{equation}
		P_{t_0}=e^{-r\left(T-t_0\right)}\left[S_{1,t_0}I_{t_0,1}^{T}-S_{2,t_0}I_{t_0,2}^{T}-KI_{t_0,3}^{T}\right],
		\label{mainresult}
	\end{equation}
	where the quantities $I_{t_0,1}^{T}$,  $I_{t_0,2}^{T}$ and $I_{t_0,3}^{T}$ are defined as:
	\begin{equation*}
		I_{t_0,1}^{T}:=\int_{d_3}^1 \tilde{q}_{1,t_0,T}\left(u\right)c_u \left(d_2\left(u\right)\right)\mbox{d}u
	\end{equation*}
	\begin{equation*}
		I_{t_0,2}^{T}:=e^{r\left(T-t_0\right)}-\int_{0}^{1} \tilde{q}_{2,t_0,T}\left(v\right)c_v\left(d_1\left(v\right)\right)\mbox{d}v
	\end{equation*}
	\begin{equation*}
		I_{t_0,3}^{T}:=\left[1- \int_{0}^{1}c_v\left(d_1\left(v\right)\right)\mbox{d}v\right]
	\end{equation*}
	and the terms $d_1\left(v\right)$, $d_2\left(u\right)$ and $d_3$  can be determined using the following three formulas
	\begin{equation*}
		d_1\left(v\right)=F_{1,t_0,T}\left(\log\left(\frac{S_{2,t_0}\tilde{q}_{2,t_0,T}\left(v\right)+K}{S_{1,t_0}}\right)\right),
	\end{equation*}
	\begin{equation*}
		d_2\left(u\right)=F_{2,t_0,T}\left(\log\left(\frac{S_{1,t_0}\tilde{q}_{1,t_0,T}\left(u\right)-K}{S_{2,t_0}}\right)\right)
	\end{equation*}
	and
	\begin{equation*}
		d_3=\left\{
		\begin{array}{lll}
			F_{1,t_0,T}\left(\log\left(\frac{K}{S_{1,t_0}}\right)\right) & \text{if} &  K>0\\
			0 & \text{if} & K\leq0
		\end{array}
		\right. .
	\end{equation*}
	All probabilities are defined under the equivalent martingale measure $\mathbb{Q}$. 
	
\end{theorem}
\begin{proof}
	We start from the expectation in \eqref{eq:ref1} that can be rewritten as:
	\begin{equation*}
		I_{t_0}^{T}=\int_{-\infty}^{+\infty} \int_{-\infty}^{+\infty}\max\left\{S_{1,t_0}e^{x_{1}}-S_{2,t_{0}}e^{x_{2}}-K;0\right\}\mbox{d}F_{t_0,T}\left(x_{1},x_{2}\right).
	\end{equation*}
	Exploiting the representation in \eqref{sklar theorem} and the twice differentiability of the copula function $C\left(u,v\right)$ we obtain the following formula for the differential of the conditional joint distribution:
	\begin{equation}
		\mbox{d}F_{t_0,T}\left(x_{1},x_{2}\right)=c\left(F_{1,t_0,T}\left(x_{1}\right),F_{2,t_0,T}\left(x_{2}\right)\right)\mbox{d}F_{1,t_0,T}\left(x_{1}\right)\mbox{d}F_{2,t_0,T}\left(x_{2}\right)
		\label{step1}
	\end{equation}
	Function $c\left(\cdot,\cdot\right)$ is defined in \eqref{eq:copula_density}.
	Substituting \eqref{step1} in the integral $I_{t_0}^{T}$, we obtain the following result:
	\scriptsize
	\begin{equation*}
		I_{t_0}^{T}=\int_{-\infty}^{+\infty} \int_{-\infty}^{+\infty}\left(S_{1,t_0}e^{x_{1}}-S_{2,t_{0}}e^{x_{2}}-K\right)^{+}c\left(F_{1,t_0,T}\left(x_{1}\right),F_{2,t_0,T}\left(x_{2}\right)\right)\mbox{d}F_{1,t_0,T}\left(x_{1}\right)\mbox{d}F_{2,t_0,T}\left(x_{2}\right),
	\end{equation*}
	\normalsize
	where $\left(x\right)^{+}=x\mathbbm{1}_{\left\{x>0\right\}}$ and $\mathbbm{1}_{\left\{x>0\right\}}$ is the indicator function that assumes value 1 if $x>0$, 0 otherwise.\newline
	Defining $u:=F_{1,t_0,T}\left(x_{1}\right)$ and $v:=F_{2,t_0,T}\left(x_{2}\right)$ and using the exponential of the quantile function, we obtain:
	\begin{equation}
		I_{t_0}^{T}=\int_0^1\int_0^1 \left(S_{1,t_0}\tilde{q}_{1,t_0,T}\left(u\right)-S_{2,t_0}\tilde{q}_{2,t_0,T}\left(v\right)-K\right)^{+}c\left(u,v\right) \mbox{d}u\mbox{d}v.
		\label{aa}
	\end{equation}
	The integral $I_{t_0}^{T}$ can be decomposed into three terms:
	\begin{equation}
		I_{t_0}^{T} = S_{1,t_0} I_{t_0,1}^{T}-S_{2,t_0}I_{t_0,2}^{T}-KI_{t_0,3}^{T}
	\end{equation}
	defined as follows:
	\begin{equation}
		I_{t_0,1}^{T}:=\int_{0}^{1}\int_{0}^{1}\tilde{q}_{1,t_0,T}\left(u\right)\mathbbm{1}_{\left\{S_{1,t_0}\tilde{q}_{1,t_0,T}\left(u\right)>S_{2,t_0}\tilde{q}_{2,t_0,T}\left(v\right)+K\right\}}c\left(u,v\right) \mbox{d}u\mbox{d}v
		\label{term1}
	\end{equation}
	\begin{equation}
		I_{t_0,2}^{T}:=\int_{0}^{1}\int_{0}^{1}\tilde{q}_{2,t_0,T}\left(v\right)\mathbbm{1}_{\left\{S_{1,t_0}\tilde{q}_{1,t_0,T}\left(u\right)>S_{2,t_0}\tilde{q}_{2,t_0,T}\left(v\right)+K\right\}}c\left(u,v\right) \mbox{d}u\mbox{d}v
		\label{term2}
	\end{equation}
	and
	\begin{equation}
		I_{t_0,3}^{T}:=\int_{0}^{1}\int_{0}^{1}K\mathbbm{1}_{\left\{S_{1,t_0}\tilde{q}_{1,t_0,T}\left(u\right)>S_{2,t_0}\tilde{q}_{2,t_0,T}\left(v\right)+K\right\}}c\left(u,v\right) \mbox{d}u\mbox{d}v.
		\label{term3}
	\end{equation}
	We introduce the following function:
	\begin{equation}
		d_1\left(v\right)=F_{1,t_0,T}\left(\log\left(\frac{S_{2,t_0}\tilde{q}_{2,t_0,T}\left(v\right)+K}{S_{1,t_0}}\right)\right),
	\end{equation}
	exploiting the strict increasingness of $\tilde{q}_{1,t_0,T}\left(u\right)$, we obtain the identity
	\begin{equation}
		\mathbbm{1}_{\left\{S_{1,t_0}\tilde{q}_{1,t_0,T}\left(u\right)>S_{2,t_0}\tilde{q}_{2,t_0,T}\left(v\right)+K\right\}}=\mathbbm{1}_{\left\{u>d_1\left(v\right)\right\}}
		\label{Ident}
	\end{equation}
	which we use, applying Fubini's theorem, to rewrite $I_{t_0,2}^{T}$ in the following way:
	\begin{eqnarray}
		I_{t_0,2}^{T}&=&\int_{0}^{1}\tilde{q}_{2,t_0,T}\left(v\right)\left[\int_{d_1\left(v\right)}^{1}c\left(u,v\right) \mbox{d}u\right]\mbox{d}v\nonumber\\
		&=&\int_{0}^{1}\tilde{q}_{2,t_0,T}\left(v\right)\left[1- c_v\left(d_1\left(v\right)\right)\right]\mbox{d}v\nonumber\\
		&=& \mathsf{E}^\mathbb{Q}_{t_0}\left[e^{X_{2,t_0,T}}\right]-\int_{0}^{1}\tilde{q}_{2,t_0,T}\left(v\right)c_v\left(d_1\left(v\right)\right)\mbox{d}v
		\label{IntI2}.
	\end{eqnarray}
	Using the martingale measure $\mathbb{Q}$, we have:
	\[
	\mathsf{E}^\mathbb{Q}_{t_0}\left[e^{X_{2,t_0,T}}\right]=e^{r\left(T-t_0\right)}
	\]
	and \eqref{IntI2} becomes:
	\begin{equation}
		I_{t_0,2}^{T} = e^{r\left(T-t_0\right)}-\int_{0}^{1}\tilde{q}_{2,t_0,T}\left(v\right)c_v\left(d_1\left(v\right)\right)\mbox{d}v.
		\label{I2}
	\end{equation}
	Using the identity in \eqref{Ident} and applying again Fubini's theorem, $I_{t_0,3}^{T}$ can be rewritten as:
	\begin{eqnarray}
		I_{t_0,3}^{T}&=&\int_{0}^{1}\left[\int_{0}^{1}\mathbbm{1}_{u>d_1\left(v\right)}c\left(u,v\right) \mbox{d}u\right]\mbox{d}v\nonumber\\
		&=& \int_{0}^{1}\left[\int_{d_1\left(v\right)}^{1}c\left(u,v\right) \mbox{d}u\right]\mbox{d}v\nonumber\\
		&=& \int_{0}^{1}\left[c_v\left(1\right)- c_v\left(d_1\left(v\right)\right)\right]\mbox{d}v\nonumber\\
		&=& \int_{0}^{1}\left[1- c_v\left(d_1\left(v\right)\right)\right]\mbox{d}v\nonumber\\
		&=& 1- \int_{0}^{1}c_v\left(d_1\left(v\right)\right)\mbox{d}v
		\label{I3}
	\end{eqnarray}
	Now we compute the integral $I_{t_0,1}^{T}$. The following identity holds true:
	\begin{equation}
		\mathbbm{1}_{\left\{S_{1,t_0}\tilde{q}_{1,t_0,T}\left(u\right)>S_{2,t_0}\tilde{q}_{2,t_0,T}\left(v\right)+K\right\}}=\mathbbm{1}_{\left\{d_2\left(u\right)>v\right\}}\mathbbm{1}_{\left\{1>u>d_3\right\}}
		\label{Ident3}
	\end{equation}
	where 
	\begin{equation}
		d_2\left(u\right):=F_{2,t_0,T}\left(\log\left(\frac{S_{1,t_0}\tilde{q}_{1,t_0,T}\left(u\right)-K}{S_{2,t_0}}\right)\right)
		\label{d2}
	\end{equation}
	and
	\begin{equation}
		d_3=\left\{
		\begin{array}{lll}
			F_{1,t_0,T}\left(\log\left(\frac{K}{S_{1,t_0}}\right)\right) & \text{if} &  K>0\\
			0 & \text{if} & K\leq0
		\end{array}
		\right.
		\label{d3}
	\end{equation}
	The right hand side of \eqref{Ident3}  comes from the fact that $d_2\left(u\right)>v$ is the condition for the payoff to be nonzero with positive probability while $1>u>d_3$ is the condition for the existence of the function $d_2\left(u\right)$. The function $d_2\left(u\right)$ is obtained using the monotonicity of $\tilde{q}_{2,t_0,T}$. \newline
	Applying Fubini's theorem, the integral $I_{t_0,1}^{T}$ becomes:
	\begin{eqnarray}
		I_{t_0,1}^{T}&=&\int_{d_3}^1 \tilde{q}_{1,t_0,T}\left(u\right)\left[\int_0^{d_2\left(u\right)}c\left(u,v\right) \mbox{d}v\right]\mbox{d}u\nonumber\\
		&=&\int_{d_3}^1 \tilde{q}_{1,t_0,T}\left(u\right)c_u\left(d_2\left(u\right)\right)\mbox{d}u
		\label{I1}
	\end{eqnarray}
	The result is obtained considering \eqref{I1}, \eqref{I2}, \eqref{I3}, \eqref{d2} and \eqref{d3}. 
\end{proof}
We observe that the pricing formula in \eqref{mainresult} can be rewritten as:
\begin{equation}
	P_{t_0}=e^{-r\left(T-t_0\right)}\left(S_{1,t_0}I_{t_0,1}^{T}+S_{2,t_0}\tilde{I}_{t_0,2}^{T}+K\tilde{I}_{t_0,3}^{T}\right)-S_{2,t_0}-Ke^{-r(T-t_0)},
	\label{mainresult2}
\end{equation}
with $\tilde{I}_{t_0,2}^{T}=e^{r\left(T-t_0\right)}-I_{t_0,2}^{T}$ and $\tilde{I}_{t_0,3}^{T}=1-I_{t_0,3}^{T}$.
Both formulas have an economic interpretation in terms of replication strategy with digital options. Indeed from \eqref{mainresult}, the spread option is replicated with the following portfolio strategy:
\begin{itemize} 
	\item A long position on a digital option with final payoff $S_{1,T}\mathbbm{1}_{\left\{S_{1,T}-S_{2,T}\geq K\right\}}$ (asset or nothing call spread option).
	\item A short position on a digital option with final payoff $S_{2,T}\mathbbm{1}_{\left\{S_{1,T}-S_{2,T}\geq K\right\}}$ (asset or nothing call spread option).
	\item A short position on a digital option with final payoff $K\mathbbm{1}_{\left\{S_{1,T}-S_{2,T}\geq K\right\}}$ (cash or nothing call spread option). 
\end{itemize}
The formula in \eqref{mainresult2} suggests an alternative replication strategy reported below:
\begin{itemize}
	\item A long position on a digital option with final payoff $S_{1,T}\mathbbm{1}_{\left\{S_{1,T}-S_{2,T}\geq K\right\}}$ (asset or nothing call spread option).
	\item A long position on a digital option with final payoff $S_{2,T}\mathbbm{1}_{\left\{S_{1,T}-S_{2,T}\leq K\right\}}$ (asset or nothing put spread option).
	\item A long position on a digital option with final payoff $K\mathbbm{1}_{\left\{S_{1,T}-S_{2,T}\leq K\right\}}$ (cash or nothing put spread option). 
	\item A short position on a portfolio composed of a unit of the $2^\text{nd}$ - asset  and a unit of a bond with maturity $T$ and face value $K$ (the sum $S_{2,t_0}+Ke^{-r(T-t_0)}$ is the amount of money received if we enter this short position at time $t_0$).
\end{itemize} 
It is worth noting that the result in \eqref{mainresult} can be applied to any copula function discussed in Section \ref{section:COP}. The routine requires the evaluation of one-dimensional integrals with bounded support that can be easily computed using several integrating techniques available in any standard programming language.

\section{Univariate Option Pricing Models}
\label{section:AGM}
In this section, we review the following models: Variance Gamma, Heston's stochastic volatility and affine Gaussian-GARCH(1,1) (HN-GARCH(1,1) hereafter). These three univariate option pricing models, widely applied in the financial industry, can be used in our formula in \eqref{mainresult} for pricing a spread option.

\subsection{Exponential L\'evy Variance Gamma Model}
We begin by reviewing the main features of the Variance Gamma model developed in \cite{madan1990variance,madan1998variance}. 
The Variance Gamma is obtained by evaluating a Brownian motion with constant drift $\theta$ and positive scale coefficient $\sigma$ at a random time that follows a Gamma L\'evy subordinator. Therefore the distribution over the interval $\left[0,t\right]$ is a Normal Variance Mean Mixture of the form:
\begin{equation*}
	X_{t} \stackrel{d}{=} \mu t + \theta V_t+ \sigma \sqrt{V_t}Z
\end{equation*}
where $\mu, \theta \in \mathbb{R}$, $V\sim \Gamma\left(\frac{t}{\nu},\frac{1}{\nu}\right)$ and $Z$ is a standard gaussian random variable independent from $V$.
Even if this model displays exponential tail behaviour, \cite{seneta2004fitting} shows that the Variance Gamma process has a good ability to reproduce the asymmetry and fat-tails in the log-returns distribution. 
\newline Under the physical measure, the asset price $\left\{S_t\right\}_{t \geq 0}$ has the following dynamics:
\begin{equation}
	S_t =S_0 e^{X_{t}}
	\label{eq:VGprice}
\end{equation}
where $S_0 \in \mathbb{R}_+$ is the initial price.\newline In this case, the market is incomplete and the choice of an equivalent martingale measure is necessary. In the numerical analysis section, we follow the approach proposed in \cite{madan1998variance} and obtain the stock price dynamics by adding a mean correction term, that is:
\begin{equation}
	S_t =S_0 e^{X_{t}-\left(\mu-r-\omega\right)t}
	\label{eq:VGprice}
\end{equation}
with $r$ the risk-free rate and correction term $\omega$ given by:
\begin{equation*}
	\omega=\frac{1}{\nu}\log \left(1-\theta \nu-\sigma^2\frac{\nu}{2}\right).
\end{equation*}
In order to determine the risk neutral distribution to apply the spread option pricing formula in \eqref{mainresult}, we adopt the following approach. The model parameters in \eqref{eq:VGprice} are estimated using the numerical procedure developed in \cite{loregian2012approximation}, which is based on the Expected Maximization algorithm provided in \cite{dempster1977maximum}. In \cite{loregian2012approximation}, the Variance Gamma parametrization is slightly different from that of \cite{madan1990variance}, however,  \cite{loregian2012approximation} shows that it is always possible to retrieve the original parametrization. The cumulative distribution of the process in \eqref{eq:VGprice} is computed numerically by means an the inversion formula proposed in \cite{gilpelaez}. 

\subsection{Heston Stochastic Volatility Model}

We consider the following dynamics for the asset price $S_t$ under the physical measure:
\begin{eqnarray}
	\mbox{d}S_t &=& \mu S_t \mbox{d}t+\sqrt{V_t}S_t \mbox{d}Z_{t,1}\nonumber\\
	\mbox{d}V_t &=& \kappa \left(\theta-V_t\right)\mbox{d}t+\chi \sqrt{V_t} \mbox{d}Z_{t,2}
	\label{HestonUnderP}
\end{eqnarray}
the drift coefficient is denoted by $\mu \in \mathbb{R}$, $\kappa,\theta,\chi $ are strictly positive scalars and $Z_{t,1}, Z_{t,2}$ are two Brownian motions with:
\begin{equation*}
	\mbox{d}\langle Z_1, Z_2 \rangle_t=\rho \mbox{d}t.
\end{equation*}
Analogously to the Variance Gamma model, the stochastic volatility process in \eqref{HestonUnderP} describes an incomplete market. Therefore, a procedure for identifying an equivalent martingale measure is required. We follow the seminal work of \cite{heston} and consider a market price of volatility risk assumed to be linear in the instantaneous variance $V_t$. This assumption preserves the structure in \eqref{HestonUnderP} under the equivalent martingale measure. Consequently, we have:
\begin{eqnarray}
	\mbox{d}S_t &=& r S_t \mbox{d}t+\sqrt{V_t}S_t \mbox{d}Z^{\mathbb{Q}}_{t,1}\nonumber\\
	\mbox{d}V_t &=& \kappa^{*} \left(\theta^{*}-V_t\right)\mbox{d}t+\chi \sqrt{V_t} \mbox{d}Z^{\mathbb{Q}}_{t,2}
	\label{HestonUnderQ}
\end{eqnarray}
with $\theta^{*}=\frac{\kappa\theta}{\kappa+\lambda}$, $\kappa^{*}=\kappa+\lambda$. Note that $\lambda$ denotes the market risk premium per unit of variance.  For the application of the result in \eqref{mainresult} we need to  determine the conditional characteristic function of the log-price in \eqref{HestonUnderQ} using the result in \cite{heston}.

\subsection{Heston-Nandi GARCH Model}
In this section we review the affine GARCH model with Normal innovations proposed in \cite{HNGARCH}, henceforth referred to as HN-GARCH. 

This specification retains the features of the GARCH introduced in \cite{Bollerslev} and overcomes the drawback related to symmetrical effects of shocks. In addition, it is able to capture simultaneously the stochastic nature of volatility and its negative correlation with spot returns, enabling quick adjustments in variance dictated by changes in market levels.

The HN-GARCH process is notably easy to apply to available data, and realized volatility is readily observable from the history of asset prices. Besides, the GARCH process defined in \cite{HNGARCH} allows option pricing in a semi-analytical closed-form formula, leveraging to a certain extent on numerical integration. Lastly, it can be proven that, as the observation interval shrinks, the results are numerically close to the continuous-time stochastic volatility model of \cite{heston}.

A fundamental assumption pertains the specification of the GARCH that models the volatility of the spot log-price as well as the process for the spot log-price itself.  
\begin{assumption}\label{assumption1}
	The spot asset price, $S_t$, follows the process below over time steps of length $\Delta$:
	\begin{subequations}
		
		\begin{equation}\label{hn garch st}
			\log(S_t)=\log(S_{t-\Delta}) + r + \lambda h_t + \sqrt{h_t}z_t,
		\end{equation} 
		\begin{equation}\label{hn garch h}
			h_t = \omega + \sum_{j=1}^{p} \beta _j h _{t-j\Delta} + \sum_{j=1}^{q} \alpha _j (z _{t-j\Delta} - \gamma \sqrt{h_{t-j\Delta}})^2.
		\end{equation}
	\end{subequations}
\end{assumption}

The continuously compounded risk-free interest rate is denoted with $r$, assumed to be constant over time, and $z_t$ is a standard Normal innovation. The process $h_t$ conditional on the information available at time $t-\Delta$ denotes the conditional variance of the log-returns. The conditional variance contributes to the mean in \eqref{hn garch st} as a risk premium, therefore the expected spot return is proportional to the variance by a parameter $\lambda$. Note that the return premium per unit of risk is proportional to the volatility term $\sqrt{h_t}$, exactly as in \cite{cox_ingersoll_ross}. 

	
	Successive analysis and implementation will focus on the ($p=1,q=1$) specification of the HN-GARCH, deemed the most parsimonious yet effective in modelling the time series. The process therefore reduces to
	\begin{equation}\label{hn garch11}
		h_t = \omega + \beta h _{t-1} + \alpha (z _{t-1} - \gamma \sqrt{h_{t-1}})^2
	\end{equation}
	The expected value of variance at period $t+1$, conditional on the information at time $t$, is given by
	\begin{equation}\label{hn garch exp var}
		\mathsf{E}_t[h_{t+1}] = \omega + \alpha + (\beta + \alpha \gamma^2)h_t
	\end{equation}
	Imposing the constraint $\beta + \alpha \gamma ^2 <1$, the first order process is mean reverting, with finite mean and variance. Further, the process will tend to a long-time unconditional variance level of
	\begin{equation}\label{lt var}
		V_L = \frac{\omega + \alpha}{1 - \beta - \alpha \gamma^2}
	\end{equation} 
	
	The process for $h_{t+\Delta}$ is fully observable at time $t$ as a function of the spot price
	\begin{equation}\label{ht process}
		h_{t+\Delta} = \omega + \beta h_t + \alpha \frac{(R_t - r - \lambda h_t - \gamma h_t)^2}{h_t}
	\end{equation}
	where $R_t$ is the log-return at time $t$ defined as $\log(S_t)-\log(S_{t-\Delta})$. The $\alpha$ parameter determines the kurtosis of the distribution, and imposes deterministic time-varying variance if set equal to zero. The parameter $\gamma$ accounts for asymmetric impact of shocks; a large negative shock increases the variance more than a large positive shock, consistently with financial theory. Indeed, given the non-negativity constraint imposed on the $\alpha$ parameter, a positive value of $\gamma$ would result in a negative correlation between the spot price and variance.
	
	Applying the measure change proposed in \cite{duan1995}, the risk neutral dynamics for the log-price and for the variance may be rewritten as follows:
	\begin{subequations}
		\begin{equation}\label{hn garch risk neutral st}
			\log(S_t)=\log(S_{t-\Delta}) + r - \frac{1}{2}h_t + \sqrt{h_t}z^\ast_t
		\end{equation} 
		\begin{equation}\label{hn garch risk neutral h}
			h_t = \omega + \beta h _{t-1} + \alpha (z^\ast _{t-1} - \gamma ^\ast \sqrt{h_{t-1}})^2
		\end{equation}
	\end{subequations}
	where
	\begin{equation*}
		z^\ast_t =z_t +(\lambda +\frac{1}{2}) \sqrt{h_t}
	\end{equation*} 
	\begin{equation*}
		\gamma ^\ast = \gamma + \lambda + \frac{1}{2}
	\end{equation*}
	\begin{equation*}
		\lambda ^\ast = -\frac{1}{2}.
	\end{equation*}
	
	Finally, \cite{HNGARCH} determine a log-affine characteristic function for the log-price, whose coefficients can be derived by means of a system of difference equations. In order to obtain the probability density function and probability distribution of the log-price under the risk neutral measure $\mathbb{Q}$ at expiry, we resort to the inverse Fourier transform introduced in \cite{gilpelaez}.

	\section{Numerical Analysis}
	\label{Num}
	
	In this section, we analyze the computational efficiency and accuracy of the result in \eqref{mainresult} for Gumbel, Clayton, Frank and Plackett copulas when the underlying follows the models described in Section \ref{section:AGM}. We perform the evaluation of the integrals in \eqref{mainresult} using two methods: midpoint and Simpson's rule. 
	Then, we compare the prices computed using our approach with Monte Carlo prices. In particular, we assess whether the prices obtained through our formula lie within the Monte Carlo confidence interval at the $95\%$ level (see \cite{BOYLE1977323} for more details).
	\newline
	For the Variance Gamma and HN-GARCH models, we use a set of parameters estimated from a dataset composed of the futures prices of the contract nearest to expiry for Brent crude oil and West Texas Intermediate (WTI) crude oil. Both contracts are quoted in U.S. Dollars. Three years of futures contract prices is gathered on a daily basis from March 1$^{\text{st}}$, 2017 to February 29$^{\text{th}}$, 2020. The dataset consists of 774 closing price records for Brent and 792 closing price records for WTI.
	
	The HN-GARCH and the Variance Gamma models are estimated through Maximum Likelihood Estimation on the series of log returns of both contracts. {In particular, the EM-algorithm has been used for the exponential Variance Gamma model, while to get the parameters for the Heston's volatility model, we use an alternative procedure based on the following steps: (i) we generate a set of European call prices using the HN-GARCH(1,1) model, then (ii) we estimate the Heston's parameters by minimizing the squared distance between HN-GARCH and Heston prices.
		
		Table \ref{tab:par_estimates_garch} shows the estimated parameters of all considered univariate models for Brent and WTI. 
		
		For Gumbel, Frank and Clayton copula, we estimate $\theta$ by maximizing the likelihood function (MLE). For these copulas, the MLE method is implemented in the Python package \texttt{pyvinecopulib} that is freely available. Plackett's $\theta$ is estimated by inverting the cross-product ratio as described in \cite{plackett}.
			\begin{table}[h]
				\noindent\makebox[\textwidth]{
					\begin{tabular}{cccccc}
						\hline\hline \\
						[\dimexpr-\normalbaselineskip+2pt]
						\multicolumn{6}{c}{\textbf{Variance Gamma}}\\
						& $\nu$ & $\theta$ & $\sigma$ & &  \\ \hline
						Brent & 0.928 & -4.848e-3 & 0.017 & &  \\
						WTI & 0.818 & -4.648e-3 & 0.018 & &   \\
						\hline\hline \\
						[\dimexpr-\normalbaselineskip+2pt]
						\multicolumn{6}{c}{\textbf{HN-GARCH(1,1)}}\\
						& $\omega$ & $\alpha$ & $\beta$ & $\gamma$ & $\lambda$      \\ \hline
						Brent &  9.124e-33   & 7.081e-6   &  0.914  &  96.505  &  -0.418  \\
						WTI   & 2.845e-4  &  7.155e-6  &  0.175  & 0.161 & -0.522         \\
						\hline\hline \\
						[\dimexpr-\normalbaselineskip+2pt]
						\multicolumn{6}{c}{\textbf{Heston}}\\
						&$\kappa^*$ & $\theta^*$ & $\chi$ & $\rho$ &  \\ \hline
						Brent & 4.475e-01 & 2.482e-02 & 2.112e-05 & 9.042e-01 &  \\
						WTI   & 1.992 & 0.089 & 5.349e-03 & 0.999 &   \\
						\hline\hline
				\end{tabular}}
				\caption{\label{tab:par_estimates_garch} Estimates of the parameters of the Variance Gamma, the HN-GARCH(1,1) and Heston's volatility.}
			\end{table}   
			Formula in \eqref{mainresult} requires the evaluation of the integrals in \eqref{I1}, \eqref{I2} and \eqref{I3}. Therefore, we compute numerically the cumulative distribution (cdf) function and the quantile function from the characteristic function. In particular, the cdf is determined using the formula proposed in \cite{gilpelaez} and approximated with the same numerical integration method used for computing the three integrals in \eqref{mainresult}. To determine the quantile function we invert the cdf using the following algorithm. For a small value of $\epsilon>0$ we identify two points $x_0$ and $x_N$ such that $F\left(x_{0}\right)<\epsilon$ and $F\left(x_{N}\right)>1-\epsilon$. Next, we construct an equally spaced grid over the interval $[x_0,x_N]$ with $N$ sub-intervals and evaluate the cdf at each point. To approximate the quantile associated to level $u$, we determine $\bar{x}_u :=\inf\left\{x_i, \ i=0,\ldots,N: \ F\left(x_i\right) \geq u\right\}$  and $\underline{x}_u :=\sup\left\{x_i, \ i=0,\ldots,N: \ F\left(x_i\right) \leq u\right\}$: (i) if $\bar{x}_u=\underline{x}_u$, the quantile of level $u$ is $\bar{x}_u$; (ii) if $F\left(\bar{x}_u\right)-F\left(\underline{x}_u\right) \leq \epsilon$, we interpolate linearly the quantile between $(F\left(\underline{x}_u\right),\underline{x}_u)$ and $(F\left(\bar{x}_u\right),\bar{x}_u)$; (iii) if $F\left(\bar{x}_u\right)-F\left(\underline{x}_u\right)> \epsilon$ we run the Newton-Raphson routine using the mid-point of $\underline{x}_u$ and $\bar{x}_u$ as a starting point (this latter part of the algorithm is not necessary when $N$ is sufficiently large).\newline
			Tables \ref{tab:precision_analytical_mid} (mid-point algorithm) and \ref{tab:precision_analytical_Simpson} (Simpson's rule) analyze the accuracy and the computational time of the formula in \eqref{mainresult} as $N$ varies.    
			\begin{table}[!h]
				\centering
				\resizebox{12cm}{!}{
					\begin{tabular}{lccccccccc}
						\hline\hline
						&\multicolumn{3}{c}{\textbf{Variance Gamma}}	& \multicolumn{3}{c}{\textbf{Heston}} & \multicolumn{3}{c}{\textbf{Heston-Nandi}}\\
						N	& 500 & 10000 & 100000 & 500 & 10000 & 100000 & 500 & 10000 & 100000 \\
						\hline\hline \\
						[\dimexpr-\normalbaselineskip+2pt]
						Strike & \multicolumn{9}{c}{\textbf{Gumbel} $\theta = 2.9$}\\
						\hline
						0	& 6.000 & 6.002 &	6.002 & 6.248 & 6.252 &	6.253 & 6.204 & 6.208 &	6.208\\
						2.5	& 3.876 & 3.877 &	3.877	& 4.294 & 4.297 &	4.297 & 4.162 & 4.169 & 4.169\\
						5	& 2.158 & 2.159 &	2.159	& 2.713 & 2.717 &	2.717	& 2.473 & 2.475 &	2.475\\
						7.5	& 1.005 & 1.005 &	1.006	& 1.567 & 1.570 & 1.570	& 1.256 & 1.254 & 1.254\\
						10	& 0.396 & 0.397 &	0.397	& 0.827 & 0.830 & 0.839	& 0.543 & 0.542 &	0.542\\
						\hline
						sec.	& 0.083 & 0.090 &	0.160 & 0.139 & 0.147 &	0.267 & 0.175 & 0.185 &	0.254\\
						\hline\hline\\
						[\dimexpr-\normalbaselineskip+2pt]
						Strike & \multicolumn{9}{c}{\textbf{Clayton} $\theta = 6.57$}\\					
						\hline
						0	  & 5.941 & 5.946 &	5.947 & 5.980 & 5.992 &	5.992 & 5.975 & 5.979 &	5.980\\
						2.5	& 3.642 & 3.647 &	3.647	& 3.875 & 3.883 &	3.883	& 3.820 & 3.823 &	3.824\\
						5	  & 1.841 & 1.845 &	1.846	& 2.448 & 2.455 &	2.456	& 2.182 & 2.186 &	2.186\\
						7.5	& 0.953 & 0.958 &	0.958	& 1.590 & 1.597 &	1.597	& 1.181 & 1.185 &	1.185\\
						10	& 0.529 & 0.533 &	0.534	& 1.061 & 1.068 &	1.068	& 0.635 & 0.639 &	0.639\\
						\hline
						sec.	& 0.091 & 0.092 &	0.156 & 0.126 & 0.135 &	0.268 & 0.155 & 0.167 &	0.267\\
						\hline\hline\\
						[\dimexpr-\normalbaselineskip+2pt]
						Strike  & \multicolumn{9}{c}{\textbf{Frank} $\theta = 25.28$}\\					
						\hline
						0    & 5.815 & 5.823 & 5.823 & 5.882 & 5.895 & 5.895 & 5.938 & 5.944 & 5.944\\
						2.5  & 3.439 & 3.446 & 3.447 & 3.648 & 3.659 & 3.659 & 3.677 & 3.681 & 3.681\\
						5    & 1.445 & 1.451 & 1.451 & 1.929 & 1.938 & 1.938 & 1.771 & 1.774 & 1.774\\
						7.5  & 0.481 & 0.486 & 0.486 & 0.987 & 0.994 & 0.994 & 0.625 & 0.628 & 0.628\\
						10   & 0.201 & 0.204 & 0.204 & 0.553 & 0.559 & 0.559 & 0.213 & 0.216 & 0.216\\
						\hline
						sec. & 0.087 & 0.091 & 0.146 & 0.133 & 0.143 & 0.223 & 0.155 & 0.166 & 0.272\\
						\hline\hline\\
						[\dimexpr-\normalbaselineskip+2pt]
						Strike  & 	\multicolumn{9}{c}{\textbf{Plackett} $\theta = 50.52$}\\					
						\hline
						0    & 6.000 & 6.002 & 6.002 & 6.148 & 6.161 & 6.161 & 6.142 & 6.149 & 6.149\\
						2.5  & 3.876 & 3.877 & 3.877 & 4.078 & 4.090 & 4.090 & 4.014 & 4.019 & 4.019\\
						5    & 2.158 & 2.159 & 2.159 & 2.482 & 2.492 & 2.492 & 2.262 & 2.266 & 2.266\\
						7.5  & 1.005 & 1.005 & 1.006 & 1.463 & 1.471 & 1.471 & 1.107 & 1.111 & 1.111\\
						10   & 0.396 & 0.397 & 0.397 & 0.879 & 0.887 & 0.886 & 0.528 & 0.533 & 0.533\\
						\hline
						sec. & 0.022 & 0.024 & 0.063 & 0.083 & 0.090 & 0.160 & 0.231 & 0.286 & 0.261\\
						\hline\hline
					\end{tabular}
				}
				\caption{\label{tab:precision_analytical_mid} Comparison of European spread call option prices with 
					time to maturity 3 months. The prices are computed using the formula in \eqref{mainresult} where 
					the integrals are evaluated by means of midpoint integration rule. 
					$N$ denotes the number of sub-intervals in the partition of the support $[0,1]$.}
			\end{table}
			
			\begin{table}[!h]
				\centering
				\resizebox{12cm}{!}{
					\begin{tabular}{lccccccccc}
						\hline\hline
						&\multicolumn{3}{c}{\textbf{Variance Gamma}}	& \multicolumn{3}{c}{\textbf{Heston}} & \multicolumn{3}{c}{\textbf{Heston-Nandi}}\\		
						N	& 500 & 10000 & 100000 & 500 & 10000 & 100000 & 500 & 10000 & 100000 \\
						\hline\hline \\
						[\dimexpr-\normalbaselineskip+2pt]
						Strike & \multicolumn{9}{c}{\textbf{Gumbel} $\theta = 2.9$}\\
						\hline
						0	  & 5.907 & 5.997 & 6.002 & 6.148 & 6.247 & 6.253 & 6.181 & 6.205 & 6.208\\
						2.5	& 3.780 & 3.872 &	3.877	& 4.188 & 4.291 &	4.297	& 4.132 & 4.167 &	4.169\\
						5	  & 2.061 & 2.154 &	2.159	& 2.607 & 2.711 &	2.717	& 2.428 & 2.473 &	2.475\\
						7.5	& 0.900 & 1.000 &	1.005	& 1.455 & 1.564 &	1.570	& 1.197 & 1.251 &	1.254\\
						10	& 0.277 & 0.391 &	0.396	& 0.709 & 0.823 &	0.830	& 0.530 & 0.537 &	0.542\\
						\hline
						sec.	& 0.083 & 0.093 &	0.163 & 0.142 & 0.173 &	0.401 & 0.281 & 0.472 &	0.552\\
						\hline\hline\\
						[\dimexpr-\normalbaselineskip+2pt]
						Strike & 	\multicolumn{9}{c}{\textbf{Clayton} $\theta = 6.57$}\\					
						\hline
						0	  & 5.914 & 5.944 &	5.947 & 5.949 & 5.989 &	5.992 & 5.977 & 5.979 &	5.980\\
						2.5	& 3.616 & 3.645 &	3.647	& 3.838 & 3.880 &	3.883	& 3.820 & 3.823 &	3.823\\
						5	  & 1.812 & 1.843 &	1.846	& 2.408 & 2.452 &	2.456	& 2.181 & 2.185 &	2.186\\
						7,5	& 0.921 & 0.955 &	0.958	& 1.546 & 1.594 &	1.597	& 1.175 & 1.184 &	1.185\\
						10	& 0.492 & 0.531 &	0.534 & 1.012 & 1.064 &	1.068 & 0.625 & 0.638 &	0.639\\
						\hline
						sec.	& 0.086 & 0.154 &	0.261	& 0.122 & 0.145 &	0.261	& 0.172 & 0.173 &	0.253\\					
						\hline\hline\\
						[\dimexpr-\normalbaselineskip+2pt]
						Strike  & 	\multicolumn{9}{c}{\textbf{Frank} $\theta = 25.28$}\\					
						\hline
						0	  & 5.787 & 5.820 &	5.823 & 5.826 & 5.890 &	5.895 & 5.977 & 5.979 &	5.980\\
						2.5	& 3.409 & 3.444 &	3.447	& 3.586 & 3.654 &	3.659	& 3.820 & 3.823 &	3.823\\
						5	  & 1.411 & 1.448 &	1.451	& 1.861 & 1.933 &	1.938 & 2.181 & 2.185 &	2.186\\
						7.5	& 0.443 & 0.483 &	0.486 & 0.910 & 0.989 &	0.994	& 1.175 & 1.184 &	1.185\\
						10	& 0.155 & 0.201 &	0.204	& 0.465 & 0.553 &	0.559	& 0.625 & 0.638 &	0.639\\
						\hline
						sec	& 0.083 & 0.090 &	0.093	& 0.139 & 0.128 &	0.206 & 0.174 & 0.173 &	0.253\\
						\hline\hline\\
						[\dimexpr-\normalbaselineskip+2pt]
						Strike  & 	\multicolumn{9}{c}{\textbf{Plackett} $\theta = 50.52$}\\					
						\hline
						0	  & 5.966 & 5.997 &	5.999 & 6.097 & 6.157 &	6.161 & 6.153 & 6.149 &	6.149\\
						2.5	& 3.729 & 3.763 &	3.765	& 4.021 & 4.086 &	4.090	& 4.020 & 4.018 &	4.019\\
						5	  & 1.908 & 1.945 &	1.948 & 2.416 & 2.487 &	2.492	& 2.263 & 2.266 &	2.266\\
						7.5	& 0.862 & 0.903 &	0.906	& 1.387 & 1.466 &	1.471	& 1.102 & 1.110 &	1.111\\
						10	& 0.400 & 0.447 &	0.450	& 0.795 & 0.881 &	0.886	& 0.517 & 0.531 &	0.533\\
						\hline
						sec.	& 0.082 & 0.090 &	0.148 & 0.139 & 0.128 &	0.206 & 0.174 & 0.201 &	0.235\\
						\hline\hline
					\end{tabular}
				}
				\caption{\label{tab:precision_analytical_Simpson} Comparison of European spread call option prices with 
					time to maturity 3 months. The prices are computed using the formula in \eqref{mainresult} where 
					the integrals are evaluated by means of Simpson's rule. $N$ denotes the number of sub-intervals in the partition of the support $[0,1]$.}
			\end{table}
			We observe that computational time increases slowly as $N$ increases; all prices seem to have a stable behaviour using both quadrature algorithms. With the largest number of grid points the Simpson's rule is slightly faster, except for the Gumbel copula.
			
			We compare our formula in \eqref{mainresult} with Monte Carlo method  and
			%
			the results are reported in Table \ref{tab:price_conf_interval} where all prices, obtained by applying \eqref{mainresult}, belong to the Monte Carlo confidence interval determined with $10^5$ simulations for all copulas and for both methods. As expected, the main advantage of our approach is the reduction of the computational time. 
			\begin{table}[!h]
				\centering
				\resizebox{12cm}{!}{
					\begin{tabular}{lcccccccccccc}
						\hline\hline \\
						[\dimexpr-\normalbaselineskip+2pt]
						&\multicolumn{4}{c}{Variance Gamma}	& \multicolumn{4}{c}{Heston} & \multicolumn{4}{c}{Heston-Nandi}\\		
						Strike & Formula &	MC &	LB &	UB & Formula &	MC &	LB &	UB & Formula &	MC &	LB &	UB \\
						\hline\hline \\
						[\dimexpr-\normalbaselineskip+2pt]
						Strike & 	\multicolumn{12}{c}{\textbf{Gumbel} $\theta=2.9$}\\
						0       & 6.002 &	5.995 & 5.970 & 6.020 &	6.253 & 6.261 & 6.228 &	6.292 & 6.208 & 6.220 & 6.192 & 6.248\\
						2.5	  & 3.877 &	3.874	& 3.851 & 3.896 &	4.297	& 4.304 & 4.276 &	4.333 & 4.169 & 4.167 & 4.143	& 4.191\\
						5	  & 2.159 &	2.155	& 2.138 & 2.173 &	2.717	& 2.724 & 2.701 &	2.747 & 2.475 & 2.477 &	2.457	& 2.496\\
						7.5	  & 1.006 &	1.005	& 0.992 & 1.018 &	1.570	& 1.575 & 1.557 &	1.593 & 1.254 & 1.256 &	1.242	& 1.271\\
						10	  & 0.397 &	0.399	& 0.391 & 0.407 &	0.830	& 0.831 & 0.817 &	0.844 & 0.542 & 0.544 & 0.534	& 0.553\\
						\hline
						MC. sec.& \multicolumn{4}{c}{0.1}	& \multicolumn{4}{c}{25.824} & \multicolumn{4}{c}{14.582}\\
						\hline\hline\\
						[\dimexpr-\normalbaselineskip+2pt]
						Strike & 	\multicolumn{12}{c}{\textbf{Clayton} $\theta= 6.57$}\\					
						\hline
						0	 & 5.947 & 5.951 & 5.925 & 5.977 &	5.992 & 5.992 & 5.956 &	6.029 & 5.980 & 5.958 & 5.930 & 5.987\\
						2.5	 & 3.647 & 3.650 & 3.626 & 3.674 &	3.883	& 3.883 & 3.848 &	3.917	& 3.824 & 3.798 & 3.772 & 3.824\\
						5	 & 1.846 & 1.849 & 1.827 & 1.870 &  2.456	& 2.453 & 2.422 &	2.483	& 2.186 & 2.178 &	2.156	& 2.200\\
						7.5	 & 0.958 & 0.964 & 0.947 & 0.981 &	1.597	& 1.590 & 1.564 &	1.617	& 1.185 & 1.173 & 1.155 & 1.190\\
						10	 & 0.534 & 0.539 & 0.526 & 0.552 &	1.068	& 1.059 & 1.036 &	1.081	& 0.639 & 0.627 & 0.614 & 0,640\\
						\hline
						-MC. sec.& \multicolumn{4}{c}{0.086}	& \multicolumn{4}{c}{32.590} & \multicolumn{4}{c}{14.006}\\
						\hline\hline\\
						[\dimexpr-\normalbaselineskip+2pt]
						Strike  & 	\multicolumn{12}{c}{\textbf{Frank} $\theta= 25.28$}\\					
						\hline
						0	& 5.823 & 5.814 &	5.795 & 5.832 & 5.895 &	5.894 & 5.867 & 5.921 &	5.944 & 5.938 & 5.918 & 5.958\\
						2.5	& 3.447 & 3.438 &	3.421	& 3.455 & 3.659 &	3.657	& 3.631 & 3.682 &	3.681	& 3.682 & 3.664 & 3.700\\
						5	& 1.451 & 1.442 &	1.427	& 1.456 & 1.938 &	1.935	& 1.912 & 1.957 &	1.774 & 1.775 & 1.760 &	1.790\\
						7.5	& 0.486 & 0.481 &	0.470	& 0.491 & 0.994 &	0.988	& 0.969 & 1.006 &	0.628	& 0.628 & 0.617 &	0,638\\
						10	& 0.204 & 0.201 &	0.193	& 0.209 & 0.559 &	0.552	& 0.537 & 0.566 &	0.216	& 0.210 & 0.204 &	0.217\\
						\hline
						MC. sec.& \multicolumn{4}{c}{0.1786}	& \multicolumn{4}{c}{25.890} & \multicolumn{4}{c}{20.361}\\
						\hline\hline\\
						[\dimexpr-\normalbaselineskip+2pt]
						Strike  & 	\multicolumn{12}{c}{\textbf{Plackett} $\theta=50.52$}\\					
						\hline
						0	& 5.999 & 5.990 &	5.965 & 6.015 & 6.161 &	6.163 & 6.129 & 6.196 &	6.149 & 6.154 & 6.127 &	6.182\\
						2.5	& 3.765 & 3.756 &	3.733	& 3.779 & 4.090 &	4.089	& 4.058 & 4.120 &	4.019	& 4.006 & 3.982 &	4.031\\
						5	& 1.948 & 1.939 &	1.920	& 1.959 & 2.492 &	2.490	& 2.463 & 2.517 &	2.266	& 2.272 & 2.251 & 2.293\\
						7.5	& 0.906 & 0.899 &	0.884	& 0.915 & 1.471 &	1.467	& 1.444 & 1.490 &	1.111	& 1.123 & 1.107 &	1.140\\
						10	& 0.450 & 0.447 &	0.435	& 0.459 & 0.886 &	0.881	& 0.862 & 0.900 &	0.533	& 0.538 & 0.526 &	0.551\\
						\hline
						MC. sec.& \multicolumn{4}{c}{0.085}	& \multicolumn{4}{c}{21.420} & \multicolumn{4}{c}{16.321}\\
						\hline\hline
					\end{tabular}
				}
				\caption{\label{tab:price_conf_interval} Comparison of 90 days European spread call option prices 
					computed using Formula \eqref{mainresult}, Monte Carlo prices (MC) and Monte 
					Carlo 95\% confidence interval. To determine the Monte Carlo upper and lower bound we 
					simulate  $10^5$ realization of the spread payoff.}
			\end{table}

\section{Copula HN-GARCH(1,1) on Log-Returns at Expiry vs Copula HN-GARCH(1,1) on Innovations: A numerical comparison}
\label{Sect:CopGARCH}

In previous works utilizing the copula GARCH pricing framework, the bivariate GARCH processes have been constructed applying a copula to generate correlated innovations (see, for example, \cite{chiou} and references therein). The procedure for generating correlated innovations is as follows. An innovation $z_{1,t}$ for the log-returns of the first asset is generated from a standard normal distribution at each time $t$. Letting $\Phi(\cdot)$ denote the standard cumulative distribution function, $u_t$ is defined as $ u_t:= \Phi(z_{1,t})$. We follow the procedure in \cite{johnson} to generate the correlated uniform random variate $v_t$. Thereafter, $v_t$ is used to compute $z_{2,t}$ using the quantile function $\Phi^{-1}(\cdot)$ for a standard normal (\textit{i.e.} $z_{2,t} = \Phi^{-1}(v_t)$) and, finally, the second asset log-returns are obtained plugging $z_{2,t}$ in the GARCH dynamics. \newline
Alternatively, following the approach we put forward in Section \ref{SOPC}, one can impose the dependence structure between the two assets through a copula function applied to log-returns at the expiry. Indeed, for the market context in Theorem \ref{MainTheorem}, it is sufficient to generate a trajectory from the HN-GARCH(1,1) model of the first asset and obtain the log-return over the entire time to maturity window $X_{1,t_0, T}$. We then compute $u_T$ as the probability transform of the log-return. The distribution, as we previously mentioned, can be obtained by means of the inversion formula \cite{gilpelaez}. Finally, via the conditional sampling method, a correlated uniform variate $v_T$ is generated and we can compute $X_{2,t_0, T}$ through the quantile function of the log-returns of the second asset.

In the sequel, we provide a comparison between these two alternative models for pricing a spread option in a GARCH context. In particular, we investigate numerically the existence of a possible relationship between the two processes' parameters. To this aim, using the formula in \eqref{mainresult} with HN-GARCH(1,1) margins, we perform a calibration exercise based on quoted spread options. We substitute the calibrated parameters in a bivariate HN-GARCH process for daily log-returns where the innovations are correlated by a Plackett copula. Finally, we measure the distance between the latter prices and those obtained by means of the pricing formula developed in Section \ref{SOPC}.

The dataset is composed of 4 trading days ranging from March 21$^{\text{st}}$ to 24$^{\text{th}}$, 2022 and is downloaded from CME Group and consists of daily closing prices for European call/put options written on the wheat-corn spread futures. 
The maturity of the spread options is the third Friday of July, while the maturity of the underlying futures is December. We verify if the options satisfy the standard no-arbitrage constraints (\textit{e.g.} Merton's constraints and convexity with respect to strike). We consider only options with moneyness $\frac{\text{strike}}{\text{today's spread}}$ ranging from 0.75 to 1.25. For each trading day we have around 30 prices. 
\begin{table}[h]
	\centering
	\begin{tabular}{lcccc}
		\hline\hline
		& March-21  & March-22  & March-23  & March-24\\
		\hline
		$\omega$	& 1.964e-32 & 1.991e-32 & 2.011e-32 & 2.976e-32\\
		$\alpha_1$	& 4.433e-07	& 4.061e-07 & 2.820e-07	& 3.425e-07\\
		$\beta_1$	& 0.932 & 0.922 & 0.759 & 0.840\\
		$\gamma_1$	& 45.097 & 44.333 & 36.619 & 35.280\\
		$\omega_2$	& 3.029e-05 & 3.036e-05 & 1.772e-05 & 2.110e-05\\
		$\alpha_2$	& 4.342e-06 & 4.381e-06 & 2.669e-07 & 9.928e-08\\
		$\beta_2$	& 0.0663	& 0.0682 &	0.1294 & 7.322e-04\\
		$\gamma_2$	& 0.524 & 0.524 &	1.010	& 1.199\\
		$\theta$	& 2.734 & 3.128 &	5.141	& 10.024\\
		\hline
		RMSE	& 0.628 &	1.812 & 4.471 &	4.202\\
		\hline\hline
	\end{tabular}
	\caption{\label{tab:calibration} Daily calibrated parameters using the pricing formula in \eqref{mainresult}. In the last row we report the value of the root of mean squared error (RMSE).}
\end{table}

We obtain the parameters in Table \ref{tab:calibration} by minimizing the squared distance between quoted prices and theoretical prices using \eqref{mainresult}. Next, we substitute the calibrated parameters of our model (hereafter Model 2) in a bivariate GARCH model for log-returns where the innovations are correlated by the Plackett copula (hereafter Model 1). 

\begin{figure}[!h]
	\centering
	\begin{subfigure}[b]{0.4\textwidth}
		\centering
		\includegraphics[width=\textwidth]{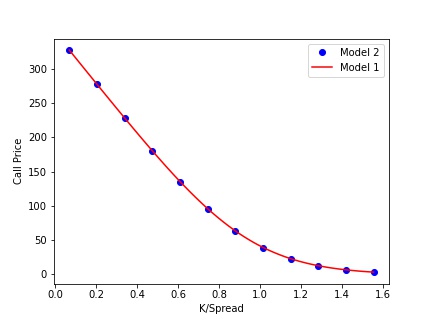}
	\end{subfigure}
	\begin{subfigure}[b]{0.4\textwidth}
		\centering
		\includegraphics[width=\textwidth]{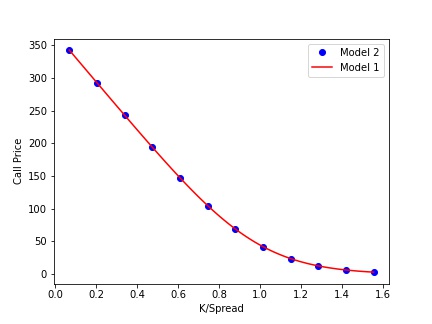}
	\end{subfigure}
	\vfill
	\begin{subfigure}[b]{0.4\textwidth}
		\centering
		\includegraphics[width=\textwidth]{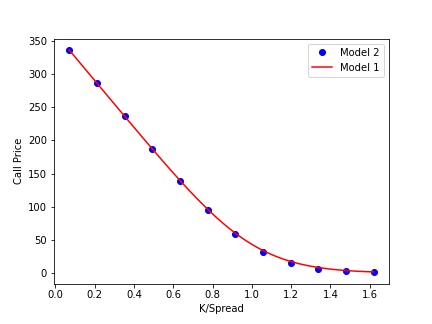}
	\end{subfigure}
	\begin{subfigure}[b]{0.4\textwidth}
		\centering
		\includegraphics[width=\textwidth]{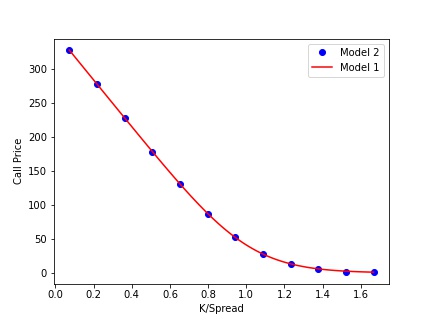}
	\end{subfigure}
	\caption{Comparison between theoretical call prices for varying strikes. To determine the prices we use the calibrated parameters in Table \ref{tab:calibration}. we denote with ``Model 1'' the bivariate GARCH model for log-returns where the innovations are obtained using the Plackett copula and "Model 2" refers to the pricing formula in \eqref{mainresult}.}
	\label{Comparison}
\end{figure}
From Figure \ref{Comparison} we observe that, feeding the parameters from Table \ref{tab:calibration} to both models, Model 1 and Model 2 yield very similar price behaviour. This observation suggests evidence of a transformation between Model 1 and Model 2 parameters. 
We denote with $\Theta_1$ and $\Theta_2$ the vectors of parameters of Model 1 and Model 2 respectively. 

Our aim is to find a linear map defined as:
\begin{equation}
	\log \left(\Theta_1\right) = A \log \left(\Theta_2\right)
	\label{eq}
\end{equation}
where the logarithm is applied to each element of $\Theta_{i}$ with $i=1,2$. Square matrix $A$ is assumed to be invertible.\newline
To estimate $A$ we perform the procedure below:
\begin{itemize}
	\item We obtain 400 randomly generated\footnote{The generated parameters satisfy the constraints imposed for the existence of a strictly positive weakly stationary process $h_t$ and the existence of a valid copula function.} observations for vector $\Theta_1$. 
	\item For each vector $\Theta_1$,  we generate 30 Monte Carlo option prices. We consider 30, 60, 90 days as maturities and, for each maturity, 10 prices with a moneyness from 0.75 to 1.25 are computed.
	\item For each simulated market, we calibrate $\Theta_2$ by minimizing the mean squared error. This results in 400 observations for vector $\Theta_2$.
	\item Using the first 300 observations we estimate matrix $A$ using the vector linear regression model. 
	\item The remaining 100 observations for $\Theta_2$ are exploited to determine $\hat{\Theta}_1$ through \eqref{eq}. Finally, we measure the performance of our procedure reporting the relative root mean squared error for each element of the vector difference $\hat{\Theta}_1-\Theta_1$. 
\end{itemize}
\begin{table}[h]
	\centering
	\resizebox{\textwidth}{!}{
		\begin{tabular}{l|ccccccccc}
			\hline\hline
			$\hat{A}$ 	& $\omega_1$ & $\alpha_1$ & $\beta_1$ & $\gamma_1$ &	$\omega_2$ & $\alpha_2$ & $\beta_2$ & $\gamma_2$ & $\theta$\\
			\hline
			$ \omega_1 $ & 0.993 &	-0.005 & -0.001 &	-0.005 & 0.004 &	0.004 & 0.004 & 0.002  &  0.003 \\
			& \scriptsize{(\texttt{1.24e-03})}   & \scriptsize{(\texttt{2.94e-03})}   & \scriptsize{(\texttt{1.46e-03})}   & \scriptsize{(\texttt{1.94e-03})}   & \scriptsize{(\texttt{1.34e-03})}   & \scriptsize{(\texttt{1.32e-03})}   & \scriptsize{(\texttt{1.38e-03})}   & \scriptsize{(\texttt{1.11e-03})}   & \scriptsize{(\texttt{2.40e-03})} \\
			$ \alpha_1 $ & 0.000 &	0.996  & -0.001 &	-0.001 & 0.003 &	0.001	& 0.001 & 0.002  & -0.001 \\
			& \scriptsize{(\texttt{1.10e-03})}   & \scriptsize{(\texttt{2.61e-03})}   & \scriptsize{(\texttt{1.29e-03})}   & \scriptsize{(\texttt{1.72e-03})}   & \scriptsize{(\texttt{1.19e-03})}   & \scriptsize{(\texttt{1.17e-03})}   & \scriptsize{(\texttt{1.23e-03})}   & \scriptsize{(\texttt{9.85e-04})}   & \scriptsize{(\texttt{2.13e-03})}\\ 
			$ \beta_1 $	 & 0.000 &	0.002  &  1.001 &	-0.002 & 0.000 &	0.002	& 0.001 & -0.001 & -0.003 \\
			& \scriptsize{(\texttt{1.07e-03})}   & \scriptsize{(\texttt{2.53e-03})}   & \scriptsize{(\texttt{1.25e-03})}   & \scriptsize{(\texttt{1.67e-03})}   & \scriptsize{(\texttt{1.15e-03})}   & \scriptsize{(\texttt{1.14e-03})}   & \scriptsize{(\texttt{1.19e-03})}   & \scriptsize{(\texttt{9.55e-04})}   & \scriptsize{(\texttt{2.07e-03})}\\ 
			$ \gamma_1 $ & 0.003 &	-0.035 & -0.006 &	0.990	 & 0.008 &	0.005	& 0.006 & 0.007  &  0.007 \\
			& \scriptsize{(\texttt{2.67e-03})}   & \scriptsize{(\texttt{6.35e-03})}   & \scriptsize{(\texttt{3.15e-03})}   & \scriptsize{(\texttt{4.20e-03})}   & \scriptsize{(\texttt{2.89e-03})}   & \scriptsize{(\texttt{2.86e-03})}   & \scriptsize{(\texttt{2.99e-03})}   & \scriptsize{(\texttt{2.40e-03})}   & \scriptsize{(\texttt{5.19e-03})}\\ 
			$ \omega_2 $ & 0.002 &	-0.004 & -0.001 &	0.002	 & 0.994 &	0.000	& -0.001 & 0.000 &  0.005 \\
			& \scriptsize{(\texttt{1.20e-03})}   & \scriptsize{(\texttt{2.84e-03})}   & \scriptsize{(\texttt{1.41e-03})}   & \scriptsize{(\texttt{1.88e-03})}   & \scriptsize{(\texttt{1.30e-03})}   & \scriptsize{(\texttt{1.28e-03})}   & \scriptsize{(\texttt{1.34e-03})}   & \scriptsize{(\texttt{1.07e-03})}   & \scriptsize{(\texttt{2.32e-03})}\\      
			$ \alpha_2 $ & 0.000 &	0.002  & 0.001  &	0.005	 & -0.001 &	0.997	& 0.000 & -0.003 & -0.002 \\
			& \scriptsize{(\texttt{1.02e-03})}   & \scriptsize{(\texttt{2.42e-03})}   & \scriptsize{(\texttt{1.20e-03})}   & \scriptsize{(\texttt{1.60e-03})}   & \scriptsize{(\texttt{1.10e-03})}   & \scriptsize{(\texttt{1.09e-03})}   & \scriptsize{(\texttt{1.14e-03})}   & \scriptsize{(\texttt{9.13e-04})}   & \scriptsize{(\texttt{1.97e-03})}\\ 
			$ \beta_2 $	& 0.001  &	0.001	& -0.001 &	-0.004 &  0.001 &	0.002	& 0.999 &	0.002 & -0.001\\
			& \scriptsize{(\texttt{1.17e-03})}   & \scriptsize{(\texttt{2.78e-03})}   & \scriptsize{(\texttt{1.38e-03})}   & \scriptsize{(\texttt{1.83e-03})}   & \scriptsize{(\texttt{1.26e-03})}   & \scriptsize{(\texttt{1.25e-03})}   & \scriptsize{(\texttt{1.31e-03})}   & \scriptsize{(\texttt{1.05e-03})}   & \scriptsize{(\texttt{2.27e-03})}\\ 
			$ \gamma_2 $ &-0.006 &	0.002	& 0.001 & 0.008 &	-0.002 & 0.000 & 0.000 & 1.000 & -0.008\\
			& \scriptsize{(\texttt{2.48e-03})}   & \scriptsize{(\texttt{5.90e-03})}   & \scriptsize{(\texttt{2.92e-03})}   & \scriptsize{(\texttt{3.90e-03})}   & \scriptsize{(\texttt{2.69e-03})}   & \scriptsize{(\texttt{2.65e-03})}   & \scriptsize{(\texttt{2.78e-03})}   & \scriptsize{(\texttt{2.23e-03})}   & \scriptsize{(\texttt{4.81e-03})}\\ 
			$ \theta $	& -0.013 &	0.019	& -0.005 &	0.004 & -0.012 &	-0.006 & -0.001 & -0.011 & 1.024\\
			& \scriptsize{(\texttt{3.46e-03})}   & \scriptsize{(\texttt{8.22e-03})}   & \scriptsize{(\texttt{4.07e-03})}   & \scriptsize{(\texttt{5.43e-03})}   & \scriptsize{(\texttt{3.75e-03})}   & \scriptsize{(\texttt{3.70e-03})}   & \scriptsize{(\texttt{3.87e-03})}   & \scriptsize{(\texttt{3.10e-03})}   & \scriptsize{(\texttt{6.71e-03})}\\
			\hline\hline
	\end{tabular}}
	\caption{\label{tab:Amatrix} Estimated coefficients of matrix $A$, the round brackets contains the corresponding standard errors.}
\end{table}
Table \ref{tab:Amatrix} reports the estimated coefficients of matrix $A$ with the corresponding standard errors. For each linear regression in \eqref{eq}, we determine the adjusted-$\text{R}^2$ that is close to 1 in all cases. From Table \ref{tab:Amatrix}, we observe that matrix $\hat{A}$ is fairly close to the identity matrix. For this reason, we report in Table \ref{tab:perRMSE} the relative mean squared error for each component under the hypotheses $A=I$ and $A=\hat{A}$.  
\begin{table}[h]
	\centering
	\resizebox{\textwidth}{!}{
		\begin{tabular}{l|ccccccccc}
			\hline\hline     
			& $\omega_1$ & $\alpha_1$ & $\beta_1$ & $\gamma_1$ & $\omega_2$ & $\alpha_2$ & $\beta_2$ & $\gamma_2$ & $\theta$\\
			\hline
			$I$ & 0.028 & 0.063 & 0.025 & 0.037 & 0.033 & 0.034 & 0.030 & 0.029 & 0.054\\
			$\hat{A}$ & 0.024 & 0.059 & 0.024 & 0.036 & 0.028 & 0.030 & 0.026 & 0.024 & 0.049\\
			\hline\hline
	\end{tabular}}
	\caption{\label{tab:perRMSE} Relative mean square error using the identity matrix $I$ and matrix $\hat{A}$ reported in Table \eqref{tab:Amatrix}.}
\end{table}

\section{Conclusion}\label{conclude}
We propose a new formula for pricing a spread option based on a generic twice differentiable copula function. The accuracy of the newly introduced formula is investigated using Monte Carlo simulations. The pricing formula depends on a single integral that can be easily computed, at least numerically, in any standard programming language. We test our formula in three option pricing models widely applied in the financial sector: Variance Gamma, Heston's stochastic volatility and HN-GARCH(1,1) models. For the latter, we also provide a numerical procedure that converts the parameters of our formula to the parameters of a bivariate HN-GARCH for log-returns where the innovations are correlated according to a copula function. This approach allows us to calibrate indirectly the latter model on the quoted option prices. 

\section*{Acknowledgment}
This work was partly supported by JST CREST Grant Number JPMJCR2115

					%
					%
					%
					%
					%
					%
					
	\newpage
	\bibliographystyle{abbrv}
	
	\bibliographystyle{abbrv}

		
		

\begin{thebibliography}{10}
	
	\bibitem{barndorff1982normal}
	O.~E. Barndorff-Nielsen, J.~Kent, and M.~S{\o}rensen.
	\newblock Normal variance-mean mixtures and z distributions.
	\newblock {\em International Statistical Review/Revue Internationale de
		Statistique}, pages 145--159, 1982.
	
	\bibitem{barndorff2001non}
	O.~E. Barndorff-Nielsen and N.~Shephard.
	\newblock Non-gaussian ornstein--uhlenbeck-based models and some of their uses
	in financial economics.
	\newblock {\em Journal of the Royal Statistical Society: Series B (Statistical
		Methodology)}, 63(2):167--241, 2001.
	
	\bibitem{bellini2020dependence}
	F.~Bellini, L.~Mercuri, and E.~Rroji.
	\newblock On the dependence structure between s\&p500, vix and implicit
	interexpectile differences.
	\newblock {\em Quantitative Finance}, 20(11):1839--1848, 2020.
	
	\bibitem{bernis2021gamma}
	G.~Bernis, R.~Brignone, S.~Scotti, and C.~Sgarra.
	\newblock A gamma ornstein--uhlenbeck model driven by a hawkes process.
	\newblock {\em Mathematics and Financial Economics}, 15(4):747--773, 2021.
	
	\bibitem{BSSpread}
	P.~Bjerksund and G.~Stensland.
	\newblock Closed form spread option valuation.
	\newblock {\em Department of Finance and Management Science, Norwegian School
		of Economics and Business Administration, Discussion Papers}, 14, 12 2006.
	
	\bibitem{Bollerslev}
	T.~Bollerslev.
	\newblock {Generalized autoregressive conditional heteroskedasticity}.
	\newblock {\em Journal of Econometrics}, 31(3):307--327, 1986.
	
	\bibitem{BOYLE1977323}
	P.~P. Boyle.
	\newblock Options: A monte carlo approach.
	\newblock {\em Journal of Financial Economics}, 4(3):323--338, 1977.
	
	\bibitem{breeden1978prices}
	D.~T. Breeden and R.~H. Litzenberger.
	\newblock Prices of state-contingent claims implicit in option prices.
	\newblock {\em Journal of business}, 51(4):621--651, 1978.
	
	\bibitem{brignone2020asian}
	R.~Brignone and C.~Sgarra.
	\newblock Asian options pricing in hawkes-type jump-diffusion models.
	\newblock {\em Annals of Finance}, 16(1):101--119, 2020.
	
	\bibitem{caldana_fusai}
	R.~Caldana and G.~Fusai.
	\newblock A general closed-form spread option pricing formula.
	\newblock {\em Journal of Banking \& Finance}, 37(12):4893--4906, 2013.
	
	\bibitem{caramona}
	R.~Carmona and V.~Durrleman.
	\newblock Pricing and hedging spread options.
	\newblock {\em SIAM Review}, 45(4):627--685, 2003.
	
	\bibitem{carr2003stochastic}
	P.~Carr, H.~Geman, D.~B. Madan, and M.~Yor.
	\newblock Stochastic volatility for l{\'e}vy processes.
	\newblock {\em Mathematical finance}, 13(3):345--382, 2003.
	
	\bibitem{carr2003finite}
	P.~Carr and L.~Wu.
	\newblock The finite moment log stable process and option pricing.
	\newblock {\em The journal of finance}, 58(2):753--777, 2003.
	
	\bibitem{chiou}
	S.~Chiou and R.~Tsay.
	\newblock A copula-based approach to option pricing and risk assesment.
	\newblock {\em Journal of Data Science}, 6:273--301, 01 2008.
	
	\bibitem{ig_garch}
	P.~Christoffersen, S.~Heston, and K.~Jacobs.
	\newblock {Option valuation with conditional skewness}.
	\newblock {\em Journal of Econometrics}, 131(1-2):253--284, 2006.
	
	\bibitem{clark1973subordinated}
	P.~K. Clark.
	\newblock A subordinated stochastic process model with finite variance for
	speculative prices.
	\newblock {\em Econometrica: journal of the Econometric Society}, pages
	135--155, 1973.
	
	\bibitem{cox_ingersoll_ross}
	J.~C. Cox, J.~E. Ingersoll, and S.~A. Ross.
	\newblock A theory of the term structure of interest rates.
	\newblock {\em Econometrica}, 53(2):385--407, 1985.
	
	\bibitem{dempster1977maximum}
	A.~P. Dempster, N.~M. Laird, and D.~B. Rubin.
	\newblock Maximum likelihood from incomplete data via the em algorithm.
	\newblock {\em Journal of the Royal Statistical Society: Series B
		(Methodological)}, 39(1):1--22, 1977.
	
	\bibitem{duan1995}
	J.~C. Duan.
	\newblock The garch option pricing model.
	\newblock {\em Mathematical Finance}, 5(1):13--32, 1995.
	
	\bibitem{duffie2000transform}
	D.~Duffie, J.~Pan, and K.~Singleton.
	\newblock Transform analysis and asset pricing for affine jump-diffusions.
	\newblock {\em Econometrica}, 68(6):1343--1376, 2000.
	
	\bibitem{eberlein2002generalized}
	E.~Eberlein and K.~Prause.
	\newblock The generalized hyperbolic model: financial derivatives and risk
	measures.
	\newblock In {\em Mathematical Finance—Bachelier Congress 2000}, pages
	245--267. Springer, 2002.
	
	\bibitem{engel}
	R.~F. Engle.
	\newblock Autoregressive conditional heteroscedasticity with estimates of the
	variance of united kingdom inflation.
	\newblock {\em Econometrica}, 50(4):987--1007, 1982.
	
	\bibitem{gilpelaez}
	J.~Gil-Pelaez.
	\newblock Note on the inversion theorem.
	\newblock {\em Biometrika}, 38(3/4):481--482, 1951.
	
	\bibitem{HHK}
	H.~Herath, P.~Kumar, and A.~Amershi.
	\newblock Crack spread option pricing with copulas.
	\newblock {\em Journal of Economics and Finance}, 37:1--22, 01 2011.
	
	\bibitem{heston}
	S.~Heston.
	\newblock A closed-form solution for options with stochastic volatility with
	applications to bond and currency options.
	\newblock {\em Review of Financial Studies}, 6:327--343, 1993.
	
	\bibitem{HNGARCH}
	S.~L. Heston and S.~Nandi.
	\newblock {A Closed-Form GARCH Option Valuation Model}.
	\newblock {\em The Review of Financial Studies}, 13(3):585--625, 06 2000.
	
	\bibitem{johnson}
	M.~E. Johnson.
	\newblock {\em Multivariate Statistical Simulation}.
	\newblock Springer Berlin Heidelberg, 2011.
	
	\bibitem{kirk}
	E.~Kirk.
	\newblock Corrleation in energy markets.
	\newblock {\em Managing Energy Price Risk}, pages 71--78, 1995.
	
	\bibitem{loregian2012approximation}
	A.~Loregian, L.~Mercuri, and E.~Rroji.
	\newblock Approximation of the variance gamma model with a finite mixture of
	normals.
	\newblock {\em Statistics \& Probability Letters}, 82(2):217--224, 2012.
	
	\bibitem{madan1998variance}
	D.~B. Madan, P.~P. Carr, and E.~C. Chang.
	\newblock The variance gamma process and option pricing.
	\newblock {\em Review of Finance}, 2(1):79--105, 1998.
	
	\bibitem{madan1990variance}
	D.~B. Madan and E.~Seneta.
	\newblock The variance gamma (vg) model for share market returns.
	\newblock {\em Journal of business}, pages 511--524, 1990.
	
	\bibitem{Malov}
	S.~V. Malov.
	\newblock {\em On Finite-Dimensional Archimedean Copulas}, pages 19--35.
	\newblock 2001.
	
	\bibitem{margrabe}
	W.~Margrabe.
	\newblock The value of an option to exchange one asset for another.
	\newblock {\em The Journal of Finance}, 33(1):177--186, 1978.
	
	\bibitem{MERCURI2008172}
	L.~Mercuri.
	\newblock Option pricing in a garch model with tempered stable innovations.
	\newblock {\em Finance Research Letters}, 5(3):172--182, 2008.
	
	\bibitem{mercuri2021finite}
	L.~Mercuri, A.~Perchiazzo, and E.~Rroji.
	\newblock Finite mixture approximation of carma (p, q) models.
	\newblock {\em SIAM Journal on Financial Mathematics}, 12(4):1416--1458, 2021.
	
	\bibitem{mercuri2018option}
	L.~Mercuri and E.~Rroji.
	\newblock Option pricing in an exponential mixedts l{\'e}vy process.
	\newblock {\em Annals of Operations Research}, 260(1):353--374, 2018.
	
	\bibitem{malliavin}
	F.~J. Mhlanga and S.~M. Kgomo.
	\newblock On the sensitivity analysis of spread options using malliavin
	calculus.
	\newblock 2021.
	
	\bibitem{muhle2012option}
	J.~Muhle-Karbe, O.~Pfaffel, and R.~Stelzer.
	\newblock Option pricing in multivariate stochastic volatility models of ou
	type.
	\newblock {\em SIAM Journal on Financial Mathematics}, 3(1):66--94, 2012.
	
	\bibitem{plackett}
	R.~L. Plackett.
	\newblock A class of bivariate distributions.
	\newblock {\em Journal of the American Statistical Association},
	60(310):516--522, 1965.
	
	\bibitem{rosenberg2003non}
	J.~V. Rosenberg.
	\newblock Non-parametric pricing of multivariate contingent claims.
	\newblock {\em The Journal of Derivatives}, 10(3):9--26, 2003.
	
	\bibitem{rroji2015mixed}
	E.~Rroji and L.~Mercuri.
	\newblock Mixed tempered stable distribution.
	\newblock {\em Quantitative Finance}, 15(9):1559--1569, 2015.
	
	\bibitem{schneider}
	L.~Schneider and B.~Tavin.
	\newblock Seasonal volatility in agricultural markets: modelling and empirical
	investigations.
	\newblock {\em Annals of Operations Research}, 2021.
	
	\bibitem{seneta2004fitting}
	E.~Seneta.
	\newblock Fitting the variance-gamma model to financial data.
	\newblock {\em Journal of Applied Probability}, 41(A):177--187, 2004.
	
	\bibitem{benth}
	T.~Sønderby~Christensen and F.~E. Benth.
	\newblock Modelling the joint behaviour of electricity prices in interconnected
	markets.
	\newblock {\em Quantitative Finance}, 20(9):1441--1456, 2020.
	
	\bibitem{vanduffel}
	J.~Van~Belle, S.~Vanduffel, and J.~Yao.
	\newblock Closed-form approximations for spread options in lévy markets.
	\newblock {\em Applied Stochastic Models in Business and Industry},
	35(3):732--746, 2019.
	
\end{thebibliography}
		
		
		
		

	\end{document}